\documentstyle[twocolumn,aps]{revtex}

\begin{document}
%\draft
%\preprint{\begin{tabular}{l}
%\end{tabular}}

\title{Embeddings of the Virasoro algebra and black hole
entropy } 
  
\author{M\'aximo Ba\~nados}

\address{Departamento de F\'{\i}sica Te\'orica, Universidad de
  Zaragoza,
  Ciudad Universitaria 50009, Zaragoza, Spain.}
 \maketitle 
 
\begin{abstract}
We consider embeddings of the Virasoro algebra into other Virasoro 
algebras with different central charges. A Virasoro algebra with
central charge $c$ (assumed to be a positive integer) and zero mode
operator $L_0$ can be embedded into another Virasoro algebra with
central charge one and zero mode operator $c L_0$. We point out that
this provides a new route to investigate the black hole entropy
problem in 2+1 dimensions. 
\end{abstract} 
~\\

Three-dimensional gravity was put forward by Deser, Jackiw and 't
Hooft \cite{djt} as an interesting toy model for gravitational
physics. It was then argued in \cite{Witten88} (see also
\cite{Achucarro-T}) that it defines a soluble and finite quantum field
theory. Questions such as what are the microscopical degrees of
freedom giving rise to the Bekenstein-Hawking entropy in three
dimensions should then have an answer. However, despite interesting
proposals \cite{Carlip,Strominger} it is clear that the
answer to this question is not yet available.  Even more, the
difficulties which arise from trying to provide a consistent quantum
description of the black hole entropy have led to the suggestion that
Einstein gravity represents a sort of thermodynamical description for
the gravitational phenomena and it thus makes no sense to attempt to
quantize it (see \cite{Martinec} and references therein). In the Loop
representation approach to quantum gravity, successful computations
for the black hole entropy, up to a numerical factor, have been
achieved. See \cite{Ashtekar-Baez-Corichi-Krasnov} and \cite{Rovelli}. 

Our main tool in analysing the three-dimensional black hole entropy
will be the discovery of Brown and Henneaux \cite{BH} that the
asymptotic symmetry group of three dimensional anti-de Sitter (adS)
space is generated by two copies of the Virasoro algebra with central
charge\cite{BH}
\begin{eqnarray}
c={3l \over 2G}
\label{c}
\end{eqnarray}
and hence, is the two-dimensional conformal group. Here $G$ is
Newton's constant, we parametrize the negative cosmological
constant as $\Lambda = -1/l^2$ and set $\hbar=1$.  In the
semiclassical regime $G\rightarrow 0$, $c$ is large. The 2+1 black
hole \cite{BHTZ} is asymptotically anti-de Sitter and thus the
conformal group acts on it in a similar form. However, globally, adS
and the black hole differ since the latter is obtained from the former
by identification of points. These identifications reduce the exact
Killing symmetries from $SO(2,2)$ to $SO(2) \times \Re$.  For this
reason, acting on the black hole, the Virasoro algebra reads\cite{CH}, 
\begin{eqnarray}
[L_n,L_m] = (n-m) L_{n+m} + \frac{c}{12} n^3 \delta_{n+m}
\label{L}
\end{eqnarray}
having a one dimensional sub-algebra generated by $L_0$. The same
holds for the other copy $\bar L_n$.  We shall call (\ref{L}) the
Ramond Virasoro algebra. The more usual Neveau-Schwarz
form of the Virasoro algebra is obtained from (\ref{L}) by shifting
the zero mode as $L_0 = L^{NS}_0-c/24$. The above convention,
appropriated to periodic boundary conditions in the spinor field, is
natural in the supergravity version of the super-conformal
algebra\cite{CH,BBCHO}. The black hole mass and angular momenta are
given in terms $L_0$ and $\bar L_0$ as, 
\begin{equation}
Ml = L_0 + \bar L_0, \ \ \ \ \ J = L_0 - \bar L_0
\end{equation}
with no added constants. With these conventions, anti-de Sitter space
has  $J=0$ and $Ml=-c/12$. For $M\geq 0$ the associated metric
represents a black hole \cite{BHTZ}, while the particle solutions
studied in \cite{Deser-Jackiw} have $-c/12 < lM <0$.   

An important open question in 2+1 adS gravity is what 
is the conformal field theory (CFT) whose energy momentum tensor
generates the $c=3l/2G$ Virasoro algebra. At the classical level (and
up to some global issues) this theory is described by a Liouville
field \cite{CHvD} (see also \cite{BBCHO}) which has two copies of the
Virasoro algebra ($L_n$, $\bar L_n$) with the correct value of $c$.
Since 2+1 gravity has no other degrees of freedom, one would expect
that the 2+1 black hole entropy is associated
to states in quantum Liouville theory with a given value of $L_0$,
$\bar L_0$. The number of such states turns out to be proportional to
the black hole area, but with a wrong power of Newton's constant $G$.
Note that this can happen because in adS$_3$ gravity there are two
length scales; the cosmological radius $l$, and the Planck length
parameter $l_p= G$. See \cite{Carlip-CT} for a first discussion
on black hole entropy and the conformal algebra, and \cite{Carlip98}
for a recent review.   

A striking observation made in \cite{Strominger} (see also
\cite{Birmingham-}) is that if the boundary CFT satisfies the
following two conditions: (i) Tr$q^{L_0}
\bar q^{\bar L_0}$ is modular invariant (with $q=\exp(2\pi i \tau)$
and Tr denotes both, the character for a given representation and the
sum over representations), and (ii) $L_0 \geq -c/24$, $\bar L_0\geq
-c/24$ (note that $L_0$ is the Ramond Virasoro operator satisfying
(\ref{L})), then the degeneracy of states for a given value of $2L_0 =
lM+J$ and $2\bar L_0=lM-J$ give
rise exactly to the Bekenstein-Hawking entropy of a
black hole of mass $M$ and angular momenta $J$.  An example of a CFT
satisfying these two conditions is a set of $c$ free bosons whose
diagonal energy momentum tensor has a central charge equal to $c$ and
satisfy both (i) and (ii).  On the contrary, Liouville theory fails to
have the right lower bound on $L_0,\bar L_0$ \cite{Kutasov-Seiberg}
and does not give the right degeneracy \cite{Carlip98}. 

The failure of Liouville theory to provide the right counting has lead
many authors to conclude that the microscopic origin of the black hole
entropy needs an additional structure (probably string theory) not
seen by the gravitational field, which would represent only an
expectation value for the true quantum fields. Specifically, in the
context of the adS/CFT conjecture
\cite{Maldacena,Witten98,Gubser-Klebanov-Polyakov}, it has
been suggested in \cite{Martinec} that the quantum CFT is related to
Liouville theory by an expression of the form $<T_{CFT}> =
T_{Liouville}$.  Incidentally, it is worth mentioning here that the
adS/CFT conjecture 
has been used to compute the central charge of $T_{CFT}$
\cite{Henningson-Skenderis,Hyun-Kim-Lee} and agreement with
the Brown-Henneaux value (\ref{c}) is found.  

Further evidence for a string theory description of the black hole was
presented in \cite{Giveon-Kutasov-Seiberg} where a string propagating
in the adS background was studied.  A formula for the spacetime
conformal generators was given in terms of the string currents. The
spacetime central charge in this approach is associated to the winding
of the worldsheet current in spacetime. Since the string theory is
unitary (for $SL(2,\Re)$ string theories see
\cite{Evans-Gaberdiel-Perry} and references therein), the
corresponding spacetime CFT is expected to give rise to the right
degeneracy. This, to our knowledge, has not not been carry out in
detail.

Whether string theory is the only solution to this problem or not is
not yet clear. However, the recent developments in the subject have
made it clear that the microscopical description of the
three-dimensional black hole entropy requires more degrees of freedom
than those arising from a naive analysis of the classical boundary
dynamics. (A counterexample to this statement is Carlip's
\cite{Carlip} original calculation which requires only an affine
$SO(2,2)$ algebra, arising in a natural way in 2+1
gravity\cite{Witten89}, plus some boundary
conditions. The main problems with that proposal seems to be the large
number of negative-norm states being counted and the physical meaning
of the boundary conditions.)     

In this paper we present a new route to attack this problem. We shall
add a set of new degrees of freedom to the classical dynamics which
upon quantization will account correctly for the Bekenstein-Hawking
entropy. These new degrees of freedom will have a simple geometrical
interpretation although their fundamental quantum origin is not yet
known to us. We shall first show our main results and then discuss
their significance and possible interpretation. 

Let $Q_n$ ($n \in Z$) be a set of operators satisfying the (Ramond)
Virasoro algebra with central charge equal to 1,
\begin{eqnarray}
[Q_n,Q_m] = (n-m) Q_{n+m} + \frac{1}{12} n^3 \delta_{n+m}.
\label{Q}
\end{eqnarray}
Irreducible and unitary representations for this algebra are known and
are uniquely classified by a highest weight state $|h>$ with conformal
weight $(Q_0 +1/24) |h> = h |h>$.  The shift $1/24$ appears here
because $Q_0$ is the Ramond Virasoro operator entering in (\ref{Q}).

For integer values of the central charge $c$, the Virasoro algebra
(\ref{L}) is a sub-algebra of (\ref{Q}) (see \cite{com} and references
therein for an extensive discussion on this point). Define the
generators $L_n$  by    
\begin{eqnarray}
L_n = {1\over c} Q_{cn} \ \ \ \ \ \ \ \ n \in Z, \ \ (c \in N).
\label{LQ}
\end{eqnarray}
For $c>1$ the $L_n$'s are a subset of the $Q_n$'s. Computing the
commutator of two $L_n$'s we obtain,
\begin{eqnarray}
[L_n,L_m] &=& \frac{1}{c^2} [Q_{cn},Q_{cm}] \nonumber\\
          &=& \frac{1}{c^2} \left(c(n-m) Q_{c(n+m)} +
          \frac{c^3n^3}{12} 
		     		  \delta_{c(n+m)} \right) \nonumber \\
		  &=& (n-m) L_{n+m} + \frac{c}{12} n^3 \delta_{n+m},
\end{eqnarray}
as desired. Note that the central
charge $c$ of (\ref{L}) needs to be an integer because otherwise
(\ref{LQ}) would not make sense. It goes without saying that if we had
started with central charge $q$ in (\ref{Q}), then the $L_n$'s defined
in (\ref{LQ}) would have central charge $qc$.  This raises an
ambiguity in the possible embeddings of (\ref{L}) into (\ref{Q}). In
our application to black hole physics we shall favour the $q=1$ case
because it gives the right degeneracy in a simple and natural form. We
expect, however, that a deeper understanding of the meaning of the
embedding at the level of the gravitational variables will provide a
better justification for this choice. 

Before going to the black hole application of this result let us
mention some consequences of the above construction \cite{com}. Start
with the algebra (\ref{Q}) with central charge $q=1/2$ (Ising model)
and choose $c=2$. The above construction means that the Ising model
has a Virasoro sub-algebra with central charge equal to one. This, of
course, can be extended to other situations and may provide unexpected
relations between the various conformal field theories\cite{Fernando}.  

The application of the above result to the black hole problem is
direct. We extend the black hole asymptotic algebra (\ref{L}) by
adding new degrees of freedom (new generators) in the way described
above such that we pass from (\ref{L}) to (\ref{Q}). We now look for
unitary representations of (\ref{Q}). The number of states ($\rho$)
for a given value of $Q_0$ grows as 
\begin{eqnarray}
\rho(Q_0) \sim \exp \left(2\pi \sqrt{Q_0/6} \right).
\label{rho}
\end{eqnarray}
Note that since the $Q$'s have central charge equal to one, this
formula is correct (for large $Q_0$). Now, since by construction $Q_0
= cL_0$ (see (\ref{LQ})) the formula (\ref{rho}) does reproduce the
right density of states when identifying the values of $L_0$ and $\bar
L_0$ with the macroscopic black hole parameters $M$ and $J$
\cite{Strominger}.  

It is instructive to see how the above mechanism applies in the
Euclidean canonical formalism.  The Euclidean black hole has the
topology of a solid torus \cite{Carlip-CT} whose boundary is a torus
with a modular parameter 
\begin{eqnarray}
\tau = {\beta \over 2\pi} \left( \Omega + {i \over l}\right),
\label{tau}
\end{eqnarray}
where $\beta$ and $\Omega$ are, respectively, the black hole
temperature and angular velocity \cite{BBO,Maldacena-S}. (This
definition of $\tau$ differs from the one used in \cite{BBO} in
the factor $2\pi$.) The gravitational partition function, under some
appropriated boundary conditions and considering only
a boundary at infinity, can then be expressed in terms of the
character\cite{CHvD,BBO}
\begin{eqnarray}
Z[\tau] = \mbox {Tr} \exp(2\pi i \tau L_0 - 2\pi i\bar \tau \bar L_0).   
\label{ZGR}
\end{eqnarray}
where $L_0$ is the zero mode Virasoro operator in (\ref{L}).  

The expected behaviour in the semiclassical Gibbons-Hawking (GH) 
approximation for $Z$ is (see \cite{BM} for a
recent discussion) 
\begin{eqnarray}
\ln Z_{GH}(\beta) = {\pi^2 l^2 \over 2G \beta},  
\label{GH}
\end{eqnarray}
where we have set $\Omega=0$ (no angular momentum) for simplicity.
This follows from evaluating the black hole free energy $-\beta
F=-\beta M + S$ with $M=r_+^2/(8Gl^2)$, $S=2\pi
r_+/(4G)$ and $\beta=2\pi l^2/r_+$. 

The question of the degeneracy of states can now be
reformulated as whether the partition function (\ref{ZGR}) reproduces
or not this semiclassical behaviour. 

If $Z[\tau]$ defined in (\ref{ZGR}) was modular invariant with 
$L_0, \bar L_0 \geq -c/24$ and $c$ given in (\ref{c}), then it
is direct to see that $Z[\tau]$ would behave exactly as (\ref{GH})
\cite{fn2}. This is nothing but the canonical version of the results
obtained in \cite{Strominger}. The trouble is that for $c>1$, either
looking at representations of the Liouville theory or the Virasoro
algebra (\ref{L}) itself, it is not possible to fulfil both
conditions. Indeed, (\ref{ZGR}) does not show the behaviour
(\ref{GH}).  The trace in (\ref{ZGR}) needs to be taken on a bigger
Hilbert space; more degrees of freedom are necessary.  

Let us take the full algebra generated by the $Q's$, of which
(\ref{L}) is a sub-algebra, and compute the trace over representations
of (\ref{Q}) with central charge one. First we use (\ref{LQ}) and
write 
\begin{equation}
Z_Q[\hat\tau] = \mbox {Tr}_Q \exp(2\pi i \hat\tau Q_0 - 2\pi
i\hat{\bar \tau} \bar Q_0).
\label{ZQ}
\end{equation}
with $\hat \tau = \tau/c$.  We have thus replaced the Virasoro
character with central charge $c$ and modular parameter $\tau$, with a
different character with central charge one and modular parameter
$\tau/c$. In symbols, 
\begin{equation}
ch(c,\tau) \rightarrow  ch(1,\hat\tau=\tau/c).
\label{ch}
\end{equation}

The character (\ref{ZQ}) can be computed exactly. After an
appropriated sum over zero modes (see below) we find \cite{fn3}
\begin{equation}
Z_Q[\hat \tau] = \frac{A}{\mbox{Im}(\hat \tau)^{1/2}
|\eta(\hat\tau)|^2} 
\label{Zex}
\end{equation}
which is invariant under modular mappings acting on $\hat \tau$. $A$
is a constant which does not depend on $\tau$. The asymptotic
behaviour of (\ref{Zex}) can be determined either by using the well
known asymptotic expansions for the Deddekind function, or by looking
at (\ref{ZQ}) and using modular invariance, as done in \cite{fn2}. In
any case, one finds 
\begin{equation}
\ln Z_Q  \sim  { i \pi \over 6 \hat\tau }     
\end{equation}
which, in terms of $\tau$, reproduces exactly the Gibbons-Hawking
approximation (\ref{GH}).  We then see a complete analogy between the
relations $Q_0=cL_0$ (microcanonical) and $\hat\tau = \tau/c$
(canonical) which encode the addition of the new degrees of freedom.
In particular, note that the semiclassical approximations $Q_0$ large
and $\hat\tau$ small are controlled by the central charge $c$, without
imposing any conditions over $L_0,\bar L_0$ or $\tau$.  This
means that the asymptotic behaviours (\ref{rho}) and (\ref{GH}) are
actually universal for all values of $M$ and $J$ since the
semiclassical condition $c>>1$ ensures both $Q_0$ large and $\hat\tau$
small\cite{Adam}.      

It is important to mention here that the exact result (\ref{Zex}) for
the partition function arose after an integration over zero modes
(see \cite{fn3}). This integration is actually not necessary
to have the right semiclassical limit. Indeed, for each representation
with conformal weight $h$, the character approaches (\ref{GH}) for
small $\beta$. We have chosen to perform the integration over $h$ in
order to find a modular invariant partition function, which is likely
to be an important property of the black hole dynamics.  Making the
integral over the conformal dimensions is also an statement on the
spectrum of the associated CFT. From the geometrical point of view, we
know that positive values of $L_0+\bar L_0$ represent black holes,
while $-c/12<L_0+\bar L_0<0$ give rise to conical singularities. The
states considered in the computation of (\ref{Zex}) have
$Q_0+1/24=h+N$ and we have integrated over all positive values of $h$.
This means $Q_0\geq -1/24$. In terms of $L_0$ this implies $L_0 \geq
-1/(24c)$. Thus, the modular invariant partition function (\ref{Zex})
does contain states corresponding to the particles studied in
\cite{Deser-Jackiw}. However, curiously, not all particle masses are
allowed but only the small region $-1/(12c) < L_0+\bar L_0 <0$. In
particular, anti-de Sitter space with $L_0+\bar L_0 = -c/12$ is not
included.

The issue of modular invariance brings into the scene another
important point. The boundary of the black hole is a torus with a
modular parameter $\tau$, and one could have expected the partition
function to be modular invariant under modular mappings acting on
$\tau$. On the contrary, we have found that $Z$ is invariant under the
modular group acting on $\hat\tau=\tau/c$.  This is not surprising
since all we have done is to use the identity $\tau L_0 = \hat\tau
Q_0$ and compute the trace over representations of $Q_n$.  (This is
summarized in (\ref{ch}).)  If correct, this scaling of the
modular parameter should have a precise meaning to be uncovered.  In
particular, one would like to know whether the addition of the new
generators, giving rise to (\ref{Q}) and $\hat\tau$  could be
understood in terms of the gravitational variables themselves, up to
some rescaling or duality transformations, or a more sophisticated
mechanism like introducing string degrees of freedom is necessary.

To summarize, we have shown that a Virasoro algebra with central
charge $c$ (c integer) can be understood as a sub-algebra of another
Virasoro algebra with central charge one. We have thus imposed the
quantization condition $c\in N$ where $c$ is the central charge
(\ref{c}), and have extended the Brown-Henneaux conformal algebra by
adding new generators. The new conformal algebra reproduces in a
natural way the semiclassical aspects of the 2+1 black hole
thermodynamics. An important open question not addressed here is the
uniqueness of this approach.  The embeddings of the Virasoro algebra
studied here are of course not unique, although it is not clear to us
that other embeddings of (\ref{L}) will give rise to the right
entropy. At any rate, a more detailed calculation of other
semiclassical quantities such as decay rates \cite{Emparan-Sachs} in
the $Q$-theory should test its uniqueness and correctness.  

The author is indebted to M. Henneaux and A. Ritz for a critical
reading of the manuscript which produced many improvements on it.
Useful discussions with M. Asorey, T. Brotz, F. Falceto, A. Gomberoff,
F. Markopoulou and M. Ortiz, and useful correspondence with M.B.
Halpern and C. Schweigert are also gratefully acknowledged. The
author thanks A. Ashtekar and L. Smolin for hospitality at the Center
for Gravitational Physics and Geometry where part of this work was
done. Financial support from CICYT (Spain) project AEN-97-1680 and the
Spanish postdoctoral program of Ministerio de Educaci\'on y Ciencia is
also acknowledged.

%-----------------------------------------


\begin{references}
\bibitem{fn2} The proof goes as follows. Let $\Omega=0$, 
then from (\ref{tau}) $\tau = i\beta/(2\pi l)$.  We first use the
symmetry of $Z$ under $\tau\rightarrow -1/\tau$ and write $Z[\beta] =
\mbox{Tr} \exp(-4\pi^2 l (L_0+\bar L_0)/\beta)$. In the limit
$\beta\rightarrow 0$ (large black holes) this sum is dominated by the
lowest eigenvalue of $L_0 + \bar L_0$ which is $-c/12$. Replacing this
value in $Z[\beta]$  and using (\ref{c}) one obtains (\ref{GH}).
\end{references}
\end{document}